\documentclass[preprint,NumberedRefs]{JASA}

\begin{document}

\title[R. Matsuda et al./Sound field decomposition based on two-stage NNs]{Sound field decomposition based on two-stage neural networks}
\author{Ryo Matsuda}
\email{matsuda.ryo.45s@st.kyoto-u.ac.jp}
\author{Makoto Otani}
\altaffiliation{ORCID: 0000-0001-8962-9304.}
\affiliation{Department of Architecture and Architectural Engineering, Graduate School of Engineering, Kyoto University, Kyoto-daigaku-katsura, Nishikyo-ku, Kyoto, 615-8540, Japan}

%\preprint{Author, JASA}	%if you want want this message to appear in upper right corner of title page

%\date{\today} 

\begin{abstract}
A method for sound field decomposition based on neural networks is proposed. The method comprises two stages: a sound field separation stage and a single-source localization stage. In the first stage, the sound pressure at microphones synthesized by multiple sources is separated into one excited by each sound source. In the second stage, the source location is obtained as a regression from the sound pressure at microphones consisting of a single sound source. The estimated location is not affected by discretization because the second stage is designed as a regression rather than a classification. Datasets are generated by simulation using Green's function, and the neural network is trained for each frequency. Numerical experiments reveal that, compared with conventional methods, the proposed method can achieve higher source-localization accuracy and higher sound-field-reconstruction accuracy.
\end{abstract}

%% pacs numbers not used

\maketitle

%  End of title page for Preprint option --------------------------------- %

\section{\label{sec:1} Introduction}
Sound field recording (i.e., recording the spatio-temporal distribution of sound pressures) is useful for better understanding the sound field through visualization and auralization of wave phenomena over a wide area. Sound field recording is an inverse problem that estimates the sound pressure at an arbitrary location in a region of interest from the sound pressure at discrete observation locations in space, such as at a single microphone array \cite{rafaely2004analysis,rafaely2004plane,park2005sound,rafaely2008spherical}, distributed microphones \cite{ueno2017sound}, or distributed microphone arrays \cite{samarasinghe2014wavefield,kaneko2021multiple}. In a three-dimensional sound field, an arbitrary sound field can be represented by a linear combination of bases such as spherical harmonics and plane waves; therefore, we consider estimating the coefficients for these bases by regression. Once those coefficients are obtained, the sound field can be reproduced for the listener using a loudspeaker array \cite{ward2001reproduction,poletti2005three,poletti2010sound,matsuda2021binaural} or a set of headphones \cite{otani2020binaural,noisternig20033d}.

When a sound field in a region is recorded, the representation of the sound field differs depending on whether the target region includes a sound source \cite{williams1999fourier,martin2006multiple}. In a region without a sound source (i.e., a region subject to the homogeneous Helmholtz equation), the sound field is represented through a straightforward spherical harmonic expansion or plane-wave expansion. However, in a region that includes a sound source, the sound field follows the inhomogeneous Helmholtz equation, which is an ill-posed problem, and cannot be directly expanded using those bases.

Therefore, a method has been proposed to decompose the sound field into a superposition of a small number of point sources by imposing sparsity on the distribution of sound sources as a constraint on the acoustical environment \cite{koyama2018sparse,koyama2019sparse}. The sparsity constraint improves the estimation accuracy even in frequencies greater than the spatial Nyquist frequency. However, these methods require discretization of candidate positions for the sound source location onto a grid in advance and thus cannot accurately estimate the sound source locations when sound sources do not exist at the pre-assumed grid points. In addition, although reducing the grid interval improves the estimation accuracy, it also leads to an increase in computational complexity and memory because of the larger number of grid points.

In contrast to the aforementioned methods that discretize a priori assumed sound source positions, sound field decomposition methods based on the reciprocity gap functional (RGF) \cite{el2011inverse} and the RGF in the spherical harmonic domain \cite{takida2020reciprocity} have been proposed as methods for gridless sound field decomposition. because these methods can directly estimate sound source positions in closed form, they are not affected by grid discretization. However, because of the effect of spatial aliasing, the frequency band with high reproduction accuracy is limited by the number of microphones and their arrangement.

Many neural-network-based methods have been proposed in the fields of sound source localization and direction-of-arrival estimation in recent years \cite{grumiaux2022survey}. Neural-network-based methods estimate the sound source positions either by classification or regression. Classification requires the prior discretization of candidate sound source positions and has the same off-grid problem encountered in the case of sparse sound field decomposition. By contrast, regression does not have the off-grid problem because the source positions can be obtained as the output of the network. In addition, the regression model has also shown better performances than classification for a single-source situation \cite{tang2019regression,perotin2019regression}. However, the performance of source localization based solely on single-frequency sound field information is unclear because most regression models have been considered in the time domain \cite{vera2018towards} or time-frequency domain \cite{subramanian2022deep,perotin2019regression,tang2019regression,adavanne2018sound} and are limited to specific sound source signals (e.g., speech).

Therefore, in the present study, we propose a sound field decomposition method that uses a regression-type neural network based solely on sound field information in a single frequency independent of the source information. The proposed method consists of two stages: In the first stage, the sound pressure at the microphones synthesized from multiple sound sources is separated into the sound pressure excited by each source. Then, in the second stage, the sound source position is obtained as a regression from the sound pressure at the microphone consisting of a single sound source. The strength of each sound source is obtained from the sound pressure in the microphone array after separation and the sound source position by linear regression. The structure of the neural networks is similar to that of the source-splitting proposed in \cite{subramanian2022deep}. However, the proposed method explicitly separates the contributions of the sound sources in the first stage using a loss function proposed in the present study. The proposed method also limits the number of sources in advance, which can impose sparsity constraints.

This paper is organized as follows: Section~\ref{sec:2} defines the problem setting of sound field decomposition based on the sparsity of the source distribution. Section~\ref{sec:3} describes our proposed method using neural networks; datasets and loss functions for training networks are also described. Section~\ref{sec:4} presents the numerical experiments and their results. Section.~\ref{sec:5} concludes this study.

\section{\label{sec:2} Sound field decomposition}
\subsection{\label{subsec:2:1} Preliminaries}
Throughout this paper, the following notations are used: matrices and vectors are denoted by uppercase and lowercase boldface, respectively. The imaginary unit $\sqrt{-1}$ is denoted by $\rm j$. Wave number is denoted by $k=\omega/c$, where $\omega$ is the angular frequency and $c$ is the sound velocity. The position vector is denoted by $\boldsymbol r=(x,y,z)\in\mathbb{R}^3$ in the Cartesian coordinate system. The time dependency is assumed as $\exp{({\rm j}\omega t)}$ and is hereafter omitted for simplicity.

\subsection{\label{subsec:2:2} Problem setting}
\begin{figure}[t]
    %% \reprintcolumnwidth is the same in preprint and reprint for
    %% ease of use for authors:
    \includegraphics[width=2in]{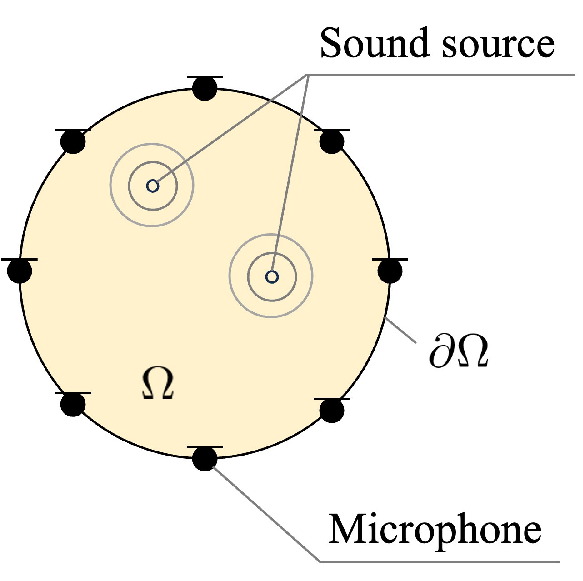}
    \caption{\label{fig:mic}{Overview of the problem setting. Sound sources exist in the target region and microphones are distributed around the region.}}
\end{figure}
Consider the reconstruction of the sound field in the region $\Omega$ in $\mathbb R^3$ including sound sources from the sound pressure measured by microphone sets $\mathcal{M}$ discretely placed on the boundary $\partial\Omega$ (Fig.~\ref{fig:mic}). Because $\Omega$ includes sources (i.e., singular points), the sound pressure in $\Omega$ satisfies the following inhomogeneous Helmholtz equation:
\begin{equation}
    (\nabla^2+k^2)p(\boldsymbol r,k)=-Q(\boldsymbol r,k).
    \label{eq:inhomo}
\end{equation}
Here, $p(\boldsymbol r,k)$ represents the sound pressure of $k$ at $\boldsymbol r\in\Omega$, $Q(\boldsymbol r,k)$ denotes the source distribution in $\Omega$, and $\nabla^2$ is the Laplace operator. The solution satisfying Eq.~(\ref{eq:inhomo}) can be expressed in terms of the volume integral of the three-dimensional free field Green's function $G(\boldsymbol r|\boldsymbol r',k)$ and $Q(\boldsymbol r',k)$ as
\begin{equation}
    p(\boldsymbol r,k)=\int_{\Omega}Q(\boldsymbol r',k)G(\boldsymbol r|\boldsymbol r',k){\rm d}\Omega,
    \label{eq:sfd2}
\end{equation}
where
\begin{equation}
    G(\boldsymbol r|\boldsymbol r',k)=\frac{\exp{(-{\rm j}k\|\boldsymbol r-\boldsymbol r'\|_2})}{4\pi\|\boldsymbol r-\boldsymbol r'\|_2}.
    \label{eq:green}
\end{equation}

If all sources in the $\Omega$ region are assumed to be point sources, Eq.~(\ref{eq:sfd2}) can be approximated as
\begin{equation}
    p(\boldsymbol r,k)\approx\sum_{s=1}^Sa_s G(\boldsymbol r|\boldsymbol r_s,k),
    \label{eq:approx}
\end{equation}
where $S$ denotes the number of sound sources and $a_s\in\mathbb C$ and $\boldsymbol r_s\in\Omega$ represent the position and amplitude of the $s$-th source, respectively. Therefore, the sound field reconstruction in $\Omega$ can be considered a sound field decomposition problem to estimate $S$, $\{a_s\}_{s\in\mathcal S}$, and $\{\boldsymbol r_s\}_{s\in\mathcal S}$ from the set of observed sound pressure $\{p(\boldsymbol r_m,k)\}_{m\in\mathcal{M}}$, where $\mathcal S$ denotes the set of the sound sources. Hereafter, the number of sound sources is assumed to be known in advance.

\section{\label{sec:3} sound field decomposition based on neural networks}
\begin{figure}[t]
    %% \reprintcolumnwidth is the same in preprint and reprint for
    %% ease of use for authors:
    \includegraphics[width=\columnwidth]{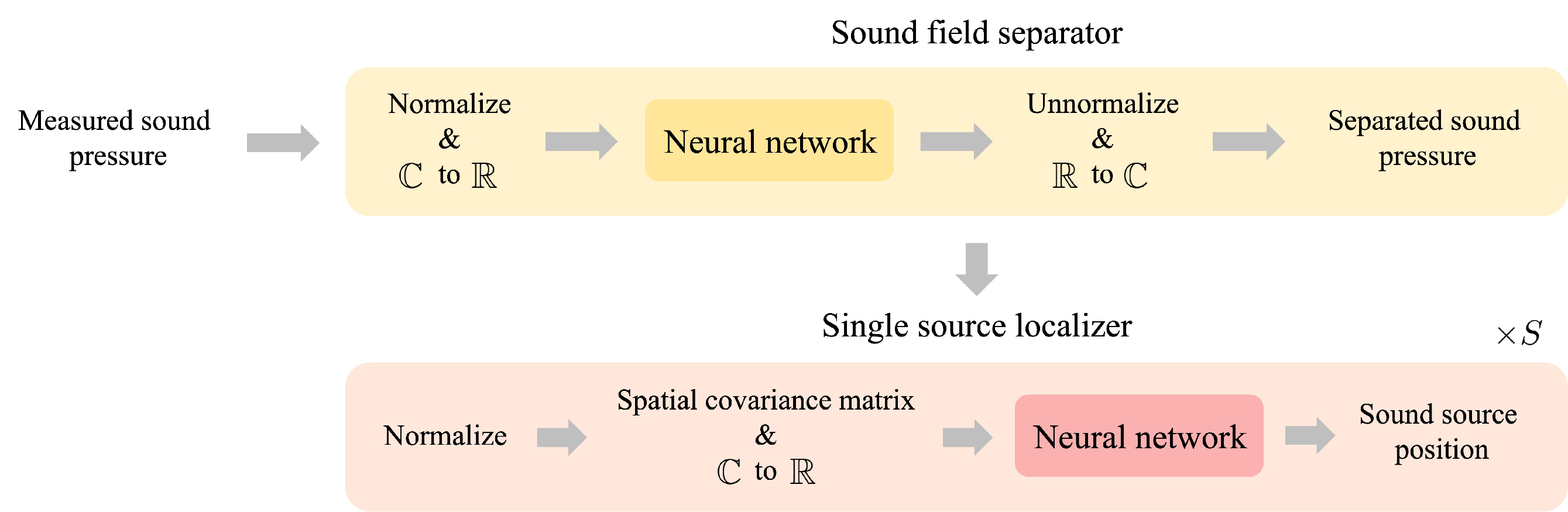}
    \caption{\label{fig:model}{Overview of the proposed sound field decomposition method based on neural networks.}}
\end{figure}

Sound field decomposition based on neural networks is proposed in this section. The proposed method consists of two stages: a sound field separator (SFS) stage and a sound source localizer (SSL) stage. Figure~\ref{fig:model} shows a schematic of the proposed model.

\subsection{\label{subsec:3:1} Model architecture}
\subsubsection{\label{subsubsec:3:1:1} Sound field separator}

\begin{figure}[t]
    %% \reprintcolumnwidth is the same in preprint and reprint for
    %% ease of use for authors:
    \includegraphics[width=0.8\columnwidth]{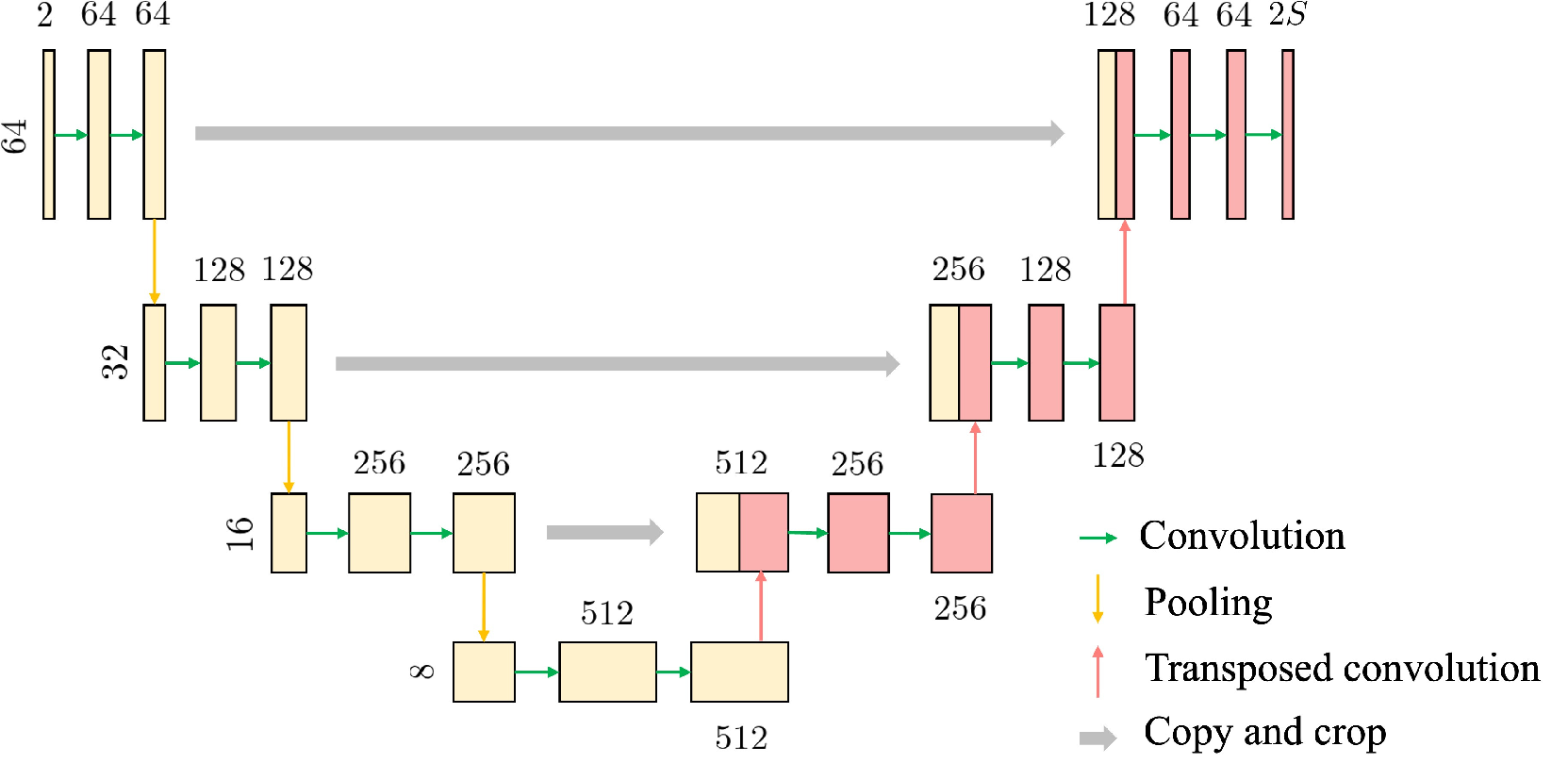}
    \caption{\label{fig:unet}{Schematic of the neural network architecture of SFS in the case of $M=64$. (Color online)}}
\end{figure}

The SFS aims to separate a sound field generated by multiple sound sources observed at the microphones into multiple sound fields generated by each source. 

The sound pressure observed at the microphones is normalized before input to the neural network as follows in order for the neural network to learn scale-independently:
\begin{equation}
    \bar{p}(\boldsymbol r_m,k)=\frac{p(\boldsymbol r_m,k)}{p_{\rm max}},
    \label{eq:normalize1}
\end{equation}
where
\begin{equation}
    p_{\rm max}=\max_{m\in\mathcal{M}}{(|p(\boldsymbol r_m,k)|)}.
\end{equation}
Here, $|\cdot|$ and $\max(\cdot)$ denote the operations of taking the absolute value and the maximum value, respectively. Because the neural network processes with real values, the complex sound pressure vector $\bar{\boldsymbol p}\in\mathbb{C^M}$, which is a column vector of $\{\bar{p}(\boldsymbol r_m,k)\}_{m\in\mathcal{M}}$, is transformed into a real-valued tensor $\bar{\boldsymbol P}^{\mathbb R}\in\mathbb{R}^{2\times M}$ as
\begin{equation}
    \begin{aligned}
    \left[\bar{\boldsymbol P}^{\mathbb R}\right]_{0,:}&=\Re\left[\bar{\boldsymbol p}^\top\right],\\
    \left[\bar{\boldsymbol P}^{\mathbb R}\right]_{1,:}&=\Im\left[\bar{\boldsymbol p}^\top\right],
    \end{aligned}
    \label{eq:sfs_input}
\end{equation}
where $\Re[\cdot]$ and $\Im[\cdot]$ represent the operations of taking the real and imaginary parts, respectively.

The neural network of the SFS is defined as a one-dimensional U-net \cite{ronneberger2015u} (Fig.~\ref{fig:unet}). Each convolution layer consists of a one-dimensional (1D) convolution followed by layer normalization, and activation, except for the final layer, which has only 1D convolution. The kernel size for convolution is 5, with stride size 1, and padding size 2. Transposed convolution is defined as 1D transposed convolution with kernel size 3, stride size 2, padding size 2, and output-padding size 2. Max pooling and rectified linear unit (ReLU) functions are used for all pooling layers and activation functions, respectively. 

The output of the neural network corresponds to a tensor of the separated sound pressure denoted by $\bar{\boldsymbol P}^{\mathbb R}_{\rm sep}\in\mathbb R^{2S\times M}$. The sound pressure vector corresponding to the $s$-th sound source, $\bar{\boldsymbol p}_{{\rm sep},s}\in\mathbb{C^M}$, is represented as
\begin{equation}
    \bar{\boldsymbol p}_{{\rm sep},s}=\left(\left[\bar{\boldsymbol P}^{\mathbb R}_{\rm sep}\right]_{2(s-1),:}+{\rm j}\left[\bar{\boldsymbol P}^{\mathbb R}_{\rm sep}\right]_{2(s-1)+1,:}\right)^\top
    \label{eq:sfs_output}
\end{equation}
and then unnormalized as
\begin{equation}
    \hat{\boldsymbol p}_{{\rm sep},s}=\bar{\boldsymbol p}_{{\rm sep},s}\times p_{\rm max}.
    \label{eq:unnormalize1}
\end{equation}

\subsubsection{\label{subsubsec:3:1:2} Single source localizer}

\begin{figure}[t]
    %% \reprintcolumnwidth is the same in preprint and reprint for
    %% ease of use for authors:
    \includegraphics[width=0.8\columnwidth]{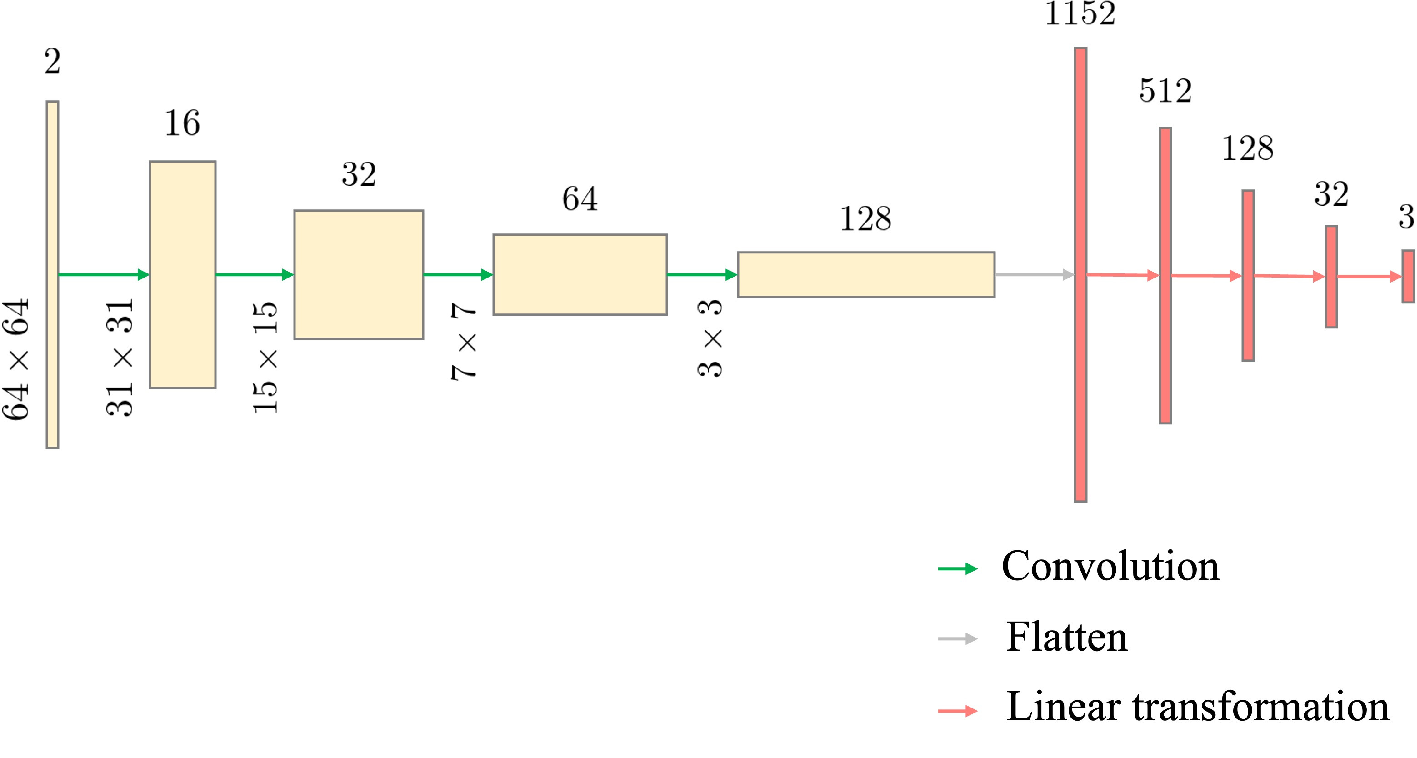}
    \caption{\label{fig:nn}{Schematic of the neural network architecture of SSL in the case of $M=64$. (Color online)}}
\end{figure}

In the SSL, the sound source position is located from the sound pressure at the microphones corresponding to each source as described in Sec.~\ref{subsubsec:3:1:1}. Therefore, the SSL is repeated $S$ times. We denote $u(\boldsymbol r_m,k)$ as the separated pressure $\hat{p}_{{\rm sep},s}(\boldsymbol r_m,k)$, and consider the $s$-th source hereafter.

The sound pressure is also normalized before the neural network for scale-independent learning as
\begin{equation}
    \bar{u}(\boldsymbol r_m,k)=\frac{u(\boldsymbol r_m,k)}{\max_{m\in\mathcal{M}}{(|u(\boldsymbol r_m,k)|)}}.
    \label{eq:normalize2}
\end{equation}
The normalized spatial covariance matrix of sound pressure vectors $\boldsymbol\Sigma=\bar{\boldsymbol u}\bar{\boldsymbol u}^{\rm H}\in\mathbb{C}^{M\times M}$ is used as the input of the neural network. Here, $\bar{\boldsymbol u}\in\mathbb{C^M}$ is a column vector of $\{\bar{u}(\boldsymbol r_m,k)\}_{m\in\mathcal{M}}$. The spatial covariance matrix is transformed into a real-valued tensor $\boldsymbol\Sigma^{\mathbb R}\in\mathbb{R}^{2\times M\times M}$ to represent it in a format compatible with the network as follows:
\begin{equation}
    \begin{aligned}
    \left[\boldsymbol\Sigma^{\mathbb R}\right]_{0,:,:}&=\Re\left[\bar{\boldsymbol u}\bar{\boldsymbol u}^{\rm H}\right],\\
    \left[\boldsymbol\Sigma^{\mathbb R}\right]_{1,:,:}&=\Im\left[\bar{\boldsymbol u}\bar{\boldsymbol u}^{\rm H}\right].
    \end{aligned}
    \label{eq:ssl_input}
\end{equation}

The neural network consists of a feature extractor composed of four convolution layers and a multilayer perceptron (MLP) composed of four linear transformation layers (Fig.~\ref{fig:nn}). Each convolution layer consists of a 2D convolution layer, layer normalization, and activation, in that order. The kernel size for convolution is $5\times 5$, the stride is 2, and the padding is 1. Each linear transformation layer except the final layer consists of a linear transformation, layer normalization, and activation, in that order. Layer normalization and activation are not used in the final layer. ReLU functions are used for all activation functions, and bias is added in all layers. The output of the neural network corresponds to the sound source position $\hat{\boldsymbol r}=(\hat{x},\hat{y},\hat{z})\in\mathbb{R}^3$ in the Cartesian coordinate system. 

The signal of the sound source can be obtained by linear regression from the estimated source position $\hat{\boldsymbol r}$ and the sound pressure at microphones $\{u(\boldsymbol r_m,k)\}_{m\in\mathcal{M}}$ as
\begin{equation}
    \left\{\begin{array}{ll}
    \hat{a}(k)=\frac{\sum_{m=1}^M\{(u(\boldsymbol r_m,k)-\mu_u)(G(\boldsymbol r_m|\hat{\boldsymbol r},k)-\mu_g)\}}{\sum_{m=1}^M(G(\boldsymbol r_m|\hat{\boldsymbol r},k)-\mu_g)^2}&{\rm if}\ S=1 \\
    \hat{\boldsymbol a} = \boldsymbol G^\dag \boldsymbol p&{\rm if}\ S>1 \\
    \end{array}\right.
    \label{eq:lin_reg1}
\end{equation}
where $\hat{\boldsymbol a}\in\mathbb{C}^S$ denotes the estimated source-signal vector; $\dag$ denotes the Moore-Penrose pseudo-inverse; $\boldsymbol G\in\mathbb{C}^{M\times S}$ denotes the transfer function matrix between the $s$-th source and the $m$-th microphone; and $\boldsymbol p\in\mathbb{C}^M$ denotes the vector of recorded sound pressure at the microphones;
\begin{equation}
    \mu_u=\frac{1}{M}\sum_{m=1}^Mu(\boldsymbol r_m,k);\quad\mu_g=\frac{1}{M}\sum_{m=1}^MG(\boldsymbol r_m|\hat{\boldsymbol r},k).
    \label{eq:lin_reg2}
\end{equation}

\subsection{\label{subsec:3:2} Dataset}
We assume that the sound field is a three-dimensional free field, that $\Omega$ is a spherical region of radius 1.0 m with the free-field condition, that microphones are located on $\partial \Omega$ with $M=64$ using a spherical {\it t}-design \cite{hardin1996mclaren}, and that the sound sources exist inside a spherical region $\Omega_{\rm S}$ of radius 0.8 m. Datasets for training the SFS and SSL are prepared separately. 

Pairs of a sound source position and simulated sound pressure at the microphones are used as the dataset for the SSL. If the sound field is assumed to be excited by a single point source, the sound pressure observed at each microphone can be obtained by
\begin{equation}
    u(\boldsymbol r_m,k)=a(k)G(\boldsymbol r_m|\boldsymbol r_{\rm src},k)+n(\boldsymbol r_m,k),
    \label{eq:dataset1}
\end{equation}
where $a(k)\in\mathbb{C}$ denotes the source signal, $\boldsymbol r_{\rm src}$ denotes the single source position, and $n(\boldsymbol r_m,k)$ denotes the noise component.

The SSL dataset comprises 10,000 pairs of $\boldsymbol r_{\rm src}$ and $\{u(\boldsymbol r_m,k)\}_{m\in\mathcal M}$ for each frequency. The positions of the sound source are randomly generated from a uniform distribution in $\Omega_S$. We use 90\% of the dataset for training and the remaining 10\% for validation. The amplitude of the sound source $|a(k)|$ is set to one, and the phase $\angle a_s(k)$ is randomly varied from batch to batch following uniform distribution ${\mathcal U}(-\pi,\pi)$ for phase-independent learning. The noise is generated as a Gaussian distribution such that the signal-to-noise ratio (SNR) is in the range of $\left[20,60\right]$ dB.

The number of sources $S$ that exist in $\Omega$ is assumed to be two for the SFS. The SFS dataset consists of pairs of the sound pressure observed at the microphones $\{p(\boldsymbol r_m,k)\}_{m\in\mathcal M}$ and the separated sound pressure corresponding to each sound source denoted by $\{p_{{\rm sep},s}(\boldsymbol r_m,k)\}_{s\in\mathcal S,m\in\mathcal M}$. The sound pressure observed at each microphone can be expressed as,
\begin{equation}
    p(\boldsymbol r_m,k)=\sum_{s=1}^S\underbrace{a_s(k)G(\boldsymbol r_m|\boldsymbol r_s,k)}_{p_{{\rm sep},s}(\boldsymbol r_m,k)}+n(\boldsymbol r_m,k).
    \label{eq:dataset2}
\end{equation}
Here, $a_s(k)\in\mathbb{C}$ denotes the signal of the $s$-th source. 

The source locations $\{\boldsymbol r_s\}_{s\in\mathcal S}$ used to generate the training dataset are selected from a random combination of 45,000 source locations used in the SSL training. Similarly, a random combination of 5,000 points from the SSL validation dataset is chosen for the source positions for validation. The source signal $\{a_s(k)\}_{s\in\mathcal S}$ is randomly varied from batch to batch following $\Re\left[a_s(k)\right]\sim{\mathcal U}(-1,1),\ \Im\left[a_s(k)\right]\sim{\mathcal U}(-1,1)$ for inter-amplitude-independent and inter-phase-independent learning. The noise is generated as a Gaussian distribution such that the SNR is in the range of $\left[20,60\right]$ dB.

\subsection{\label{subsec:3:3} Loss function and training procedure}
The mean squared error (MSE) with respect to the source position is used as a loss function of the SSL:
\begin{equation}
    \mathcal{L}_{\rm SSL}=\frac{1}{3}\|\boldsymbol r_{\rm src}-\hat{\boldsymbol r}_{\rm src}\|^2_2.
    \label{eq:loss1}
\end{equation}
Here, $\hat{\boldsymbol r}_{\rm src}$ denotes the estimated source position in each dataset, respectively. The Adam optimizer \cite{2015-kingma} with a learning rate of $5\times10^{-4}$ is used for training, and the batch size is 100. The model is trained during 1,000 epochs.

As a loss function of the SFS, we propose the permutation-invariant MSE \cite{yu2017permutation} with respect to the separated sound pressure. The loss function in the case of $S=2$ is defined as,
\begin{equation}
    \begin{aligned}
    \mathcal{L}_{\rm SFS}=\frac{1}{S}\min\Bigl({\rm MSE}_{11}+{\rm MSE}_{22},{\rm MSE}_{12}+{\rm MSE}_{21}\Bigr),
    \end{aligned}
    \label{eq:loss2}
\end{equation}
where 
\begin{equation}
    {\rm MSE}_{ij}=\frac{1}{M}\left(\sum_M|p_{{\rm sep},i}(\boldsymbol r_m,k)-\hat{p}_{{\rm sep},j}(\boldsymbol r_m,k)|^2\right).
    \label{eq:loss3}
\end{equation}
The Adam optimizer with a learning rate of $1\times10^{-3}$ is used for training, and the batch size is 100. The model is trained during 10,000 epochs.

\newpage
\section{\label{sec:4} Numerical experiments}
Numerical simulations were conducted to compare the performance of the proposed method with that of conventional methods (i.e., the sparse sound field decomposition method \cite{koyama2018sparse} and the spherical-harmonic-domain RGF \cite{takida2020reciprocity}). Hereafter, these methods are denoted as \textbf{Proposed}, \textbf{Sparse}, and \textbf{SHD-RGF}, respectively.

The arrangement of microphones was the same as that defined in Sec.~\ref{subsec:3:3}. Numerical simulations were performed for one and two sound sources, respectively.

To validate the effectiveness of SFS in the case of two sources, we used a neural-network-based model in which the output layer of the SSL model was changed to two source positions; this model was used as a baseline model, hereafter referred to as \textbf{Baseline}. The MSE of the permutation-invariant source positions was used as a loss function to learn the baseline, defined as
\begin{equation}
    \mathcal{L}_{\rm base}=\frac{1}{3S}\min\Bigl(\|\boldsymbol r_1-\hat{\boldsymbol r}_1\|^2_2+\|\boldsymbol r_2-\hat{\boldsymbol r}_2\|^2_2,\|\boldsymbol r_1-\hat{\boldsymbol r}_2\|^2_2+\|\boldsymbol r_2-\hat{\boldsymbol r}_1\|^2_2\Bigr).
    \label{eq:loss4}
\end{equation}
The dataset and noise addition used for training \textbf{Baseline} were the same as those used for training the SFS neural network. The Adam optimizer with a learning rate of $5\times10^{-4}$ was used for optimization. The model was trained during 10,000 epochs with a batch size of 100.

In addition, a neural network model with the same layer as proposed for the SFS trained by Eq.~(\ref{eq:loss4}) was used to validate the effectiveness of loss function Eq.~(\ref{eq:loss2}) in the case of two sources. The SSL of the model was pre-trained, and the weights of each SSL layer were fixed when training the SFS. The model was optimized by the Adam optimizer with a learning rate of $1\times10^{-3}$ and trained during 10,000 epochs with a batch size of 100. Hereafter, the model is denoted by \textbf{Proposed} ($\mathcal{L}_{\rm base}$).

In \textbf{Sparse}, it is necessary to discretize $\Omega$ in advance in order to set up the candidate source positions. In this experiment, $\Omega$ was discretized in the $x,y,z$ directions into grids with $\delta$ intervals and the grid points were used as source-position candidates. Therefore, we discretized the source-included region $\Omega_s$ by $\delta=0.1$ m and $\delta=0.2$ m; the total number of candidate points was 2,109 and 257, respectively. Hereafter, they are denoted by \textbf{Sparse}~($\delta=0.1$) and \textbf{Sparse}~($\delta=0.2$), respectively. The OMP \cite{pati1993orthogonal,tropp2007signal} algorithm was used for sparse decomposition.

Because \textbf{SHD-RGF} requires truncation of the order of the spherical harmonic expansion, three truncation orders (i.e., 5, 6, and 7) were used; they are denoted by \textbf{SHD-RGF}~($N=5$), \textbf{SHD-RGF}~($N=6$), and \textbf{SHD-RGF}~($N=7$), respectively.

In this experiment, we compare and evaluate each method in terms of the accuracy of sound source localization and sound field reconstruction. To evaluate the accuracy of the sound source localization, we define the root-means-square error (RMSE) as 
\begin{equation}
    {\rm RMSE}=
    \left\{
    \begin{array}{ll}
    \sqrt{\|\boldsymbol r_1-\hat{\boldsymbol r}_1\|^2_2} &{\rm if}\ S=1 \\
    \sqrt{\frac{1}{S}\min\Bigl(\|\boldsymbol r_1-\hat{\boldsymbol r}_1\|^2_2+\|\boldsymbol r_2-\hat{\boldsymbol r}_2\|^2_2,\|\boldsymbol r_1-\hat{\boldsymbol r}_2\|^2_2+\|\boldsymbol r_2-\hat{\boldsymbol r}_1\|^2_2\Bigr)} &{\rm if}\ S=2.
    \end{array}
    \right.
    \label{eq:rmse}
\end{equation}
To evaluate the accuracy of the sound field reconstruction, we define the signal-to-distortion ratio (SDR) as
\begin{equation}
    {\rm SDR}=10\log_{10}\frac{\int_{\Omega}|p_{\rm rec}(\boldsymbol r,k)-p(\boldsymbol r,k)|^2{\rm d}\boldsymbol r}{\int_{\Omega}|p(\boldsymbol r,k)|^2{\rm d}\boldsymbol r}\ ({\rm dB}).
    \label{eq:sdr}
\end{equation}
$\Omega$ was discretized at 0.1 m intervals to calculate the integral in Eq.~(\ref{eq:sdr}).

\subsection{\label{subsec:4:1} Training results}
All neural networks were trained using a single GPU (GeForce RTX 3090, NVIDIA).

\begin{figure}[t]
    \includegraphics[width=\columnwidth]{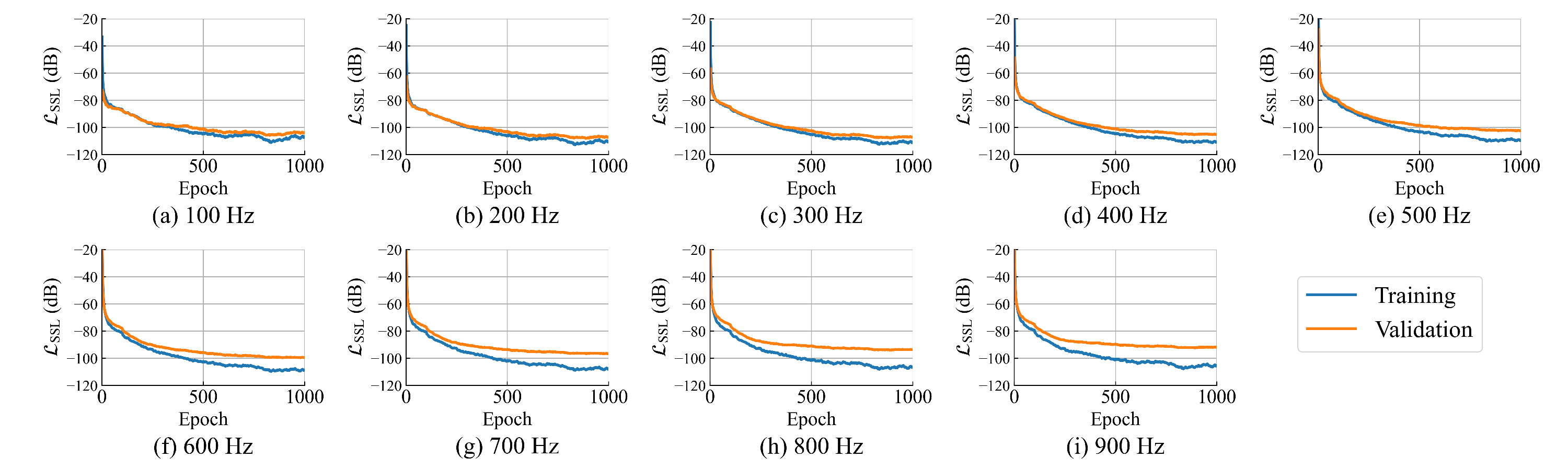}
    \caption{\label{fig:loss_ssl}{Training and validation loss of SSL plotted against the epoch number at each frequency. (Color online)}}
\end{figure}

Figure~\ref{fig:loss_ssl} shows the training and validation loss of the SSL as a function of the epoch number. The training loss was found to decrease with each additional epoch at all frequencies. However, the difference between the validation loss and the training loss increased slightly with increasing frequency. The computation time for SSL training was \verb|~|0.7 h for each frequency.

\begin{figure}[]
    \includegraphics[width=\columnwidth]{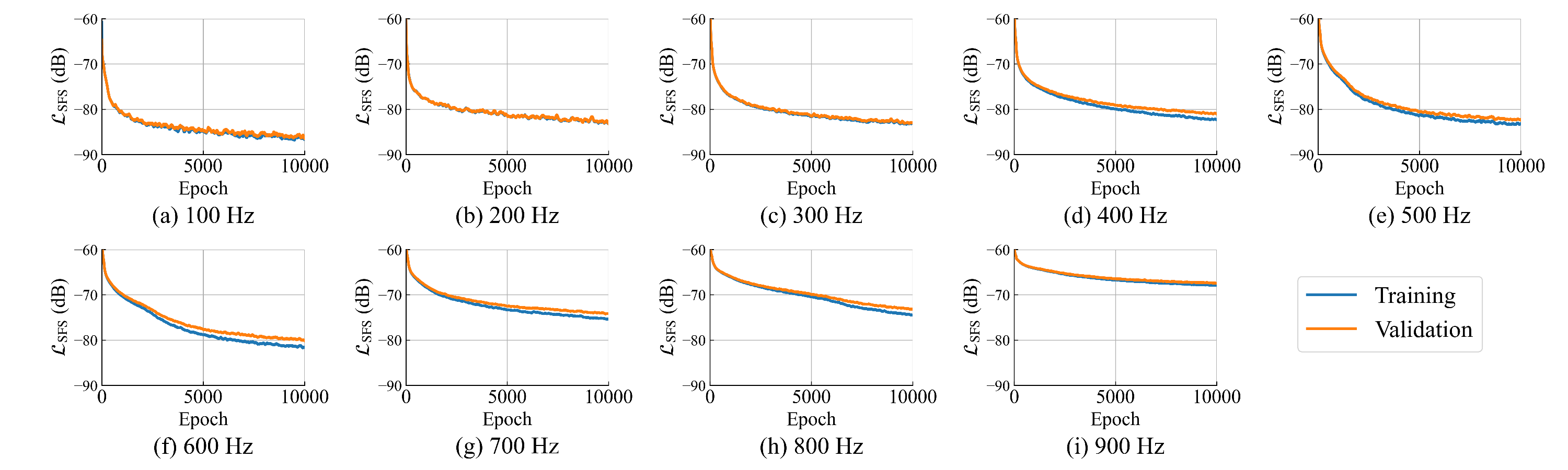}
    \caption{\label{fig:loss_sfs}{Training and validation loss of SFS for \textbf{Proposed} plotted against the epoch number at each frequency. (Color online)}}
\end{figure}

Figure~\ref{fig:loss_sfs} shows the training and validation loss of the SFS for \textbf{Proposed} as a function of the epoch number. Unlike the SSL learning, little difference was observed between the training loss and the validation loss at all frequencies. However, the loss converged to a larger value as the frequency increased. The computation time for SFS training in \textbf{Proposed} was about \verb|~|16.7 h for each frequency.

\begin{figure}[]
    \includegraphics[width=\columnwidth]{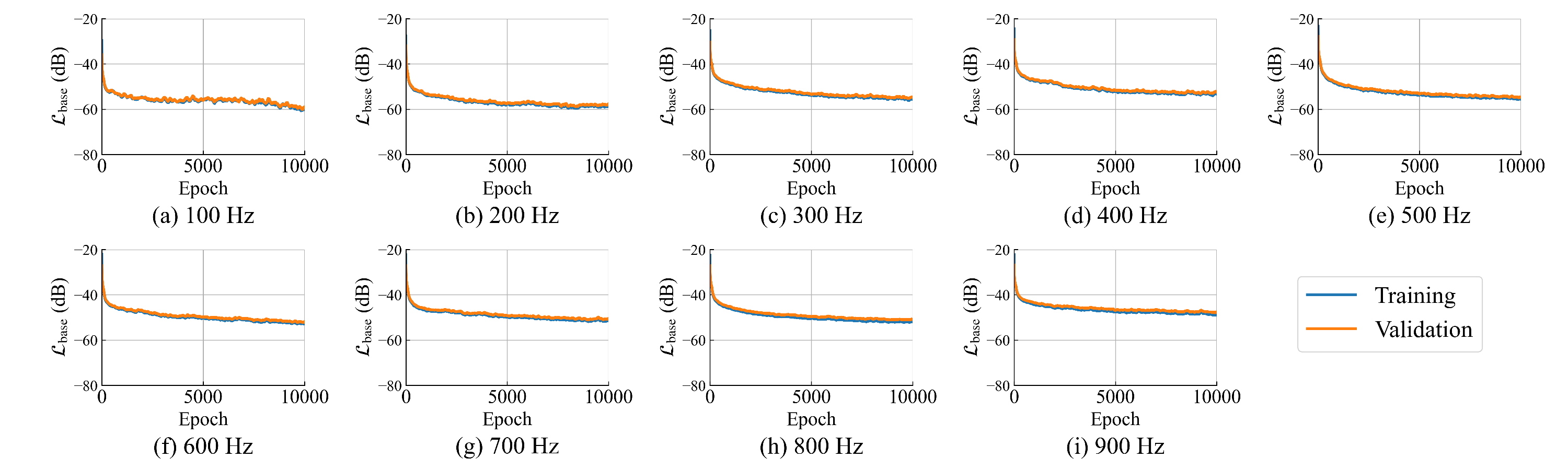}
    \caption{\label{fig:loss_base1}{Training and validation loss of \textbf{Baseline} plotted against the epoch number at each frequency. (Color online)}}
\end{figure}

Figure~\ref{fig:loss_base1} shows the training and validation loss of \textbf{Baseline} as a function of the epoch number. Although the converged loss values were dependent on the frequency, the training loss and validation loss converged similarly for all of the investigated frequencies. The computation time for training in \textbf{Baseline} was about \verb|~|14.0 h for each frequency.

\begin{figure}[]
    \includegraphics[width=\columnwidth]{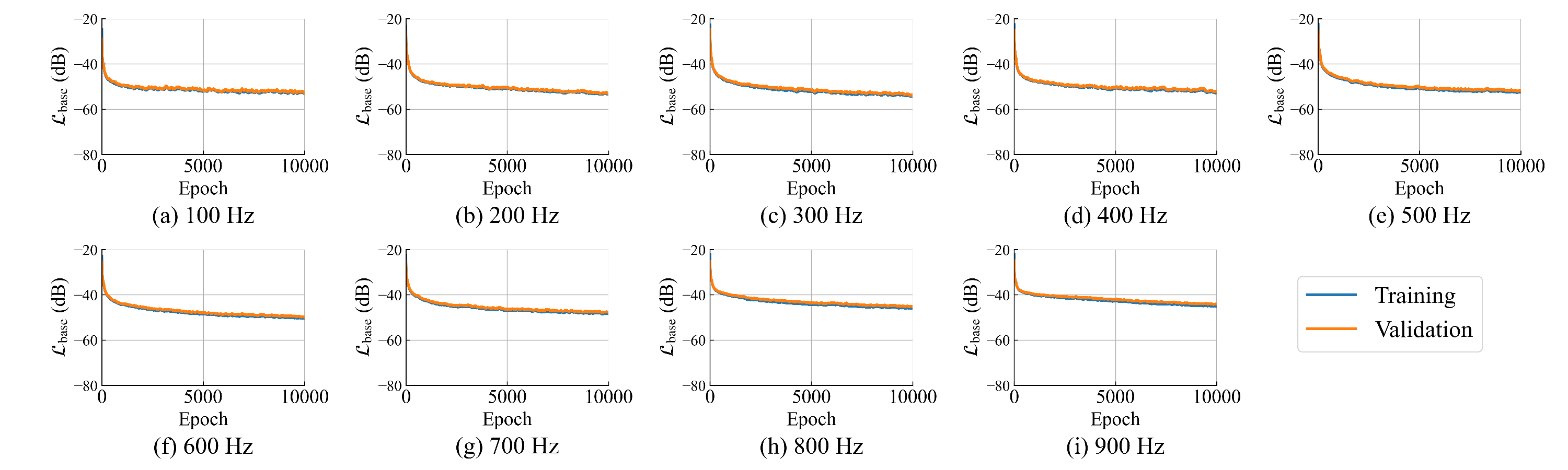}
    \caption{\label{fig:loss_base2}{Training and validation loss of SFS for \textbf{Proposed}~($\mathcal{L}_{\rm base}$) plotted against the epoch number at each frequency. (Color online)}}
\end{figure}

Figure~\ref{fig:loss_base2} shows the training and validation loss of \textbf{Proposed}~($\mathcal{L}_{\rm base}$) as a function of the epoch number. The learning trend was similar to that of \textbf{Baseline}; however, the loss value was slightly greater. The computation time for training SFS in \textbf{Proposed} ($\mathcal{L}_{\rm base}$) was about \verb|~|20.7 h for each frequency. 

\clearpage

\subsection{\label{subsec:4:2} Experiments for a single source}
In the case of a single sound source, \textbf{Proposed} consists of the SSL only. The source position sets used in this simulation were the entire SSL validation dataset described in Sec.~\ref{subsec:3:3}. The amplitude of the source signal was set to one, and the phase was randomly chosen from ${\mathcal U}(-\pi,\pi)$ for each condition. We compared each method with an SNR of 40 dB and 20 dB. The results were averaged for all conditions.

\begin{figure}[t]
    \includegraphics[width=\columnwidth]{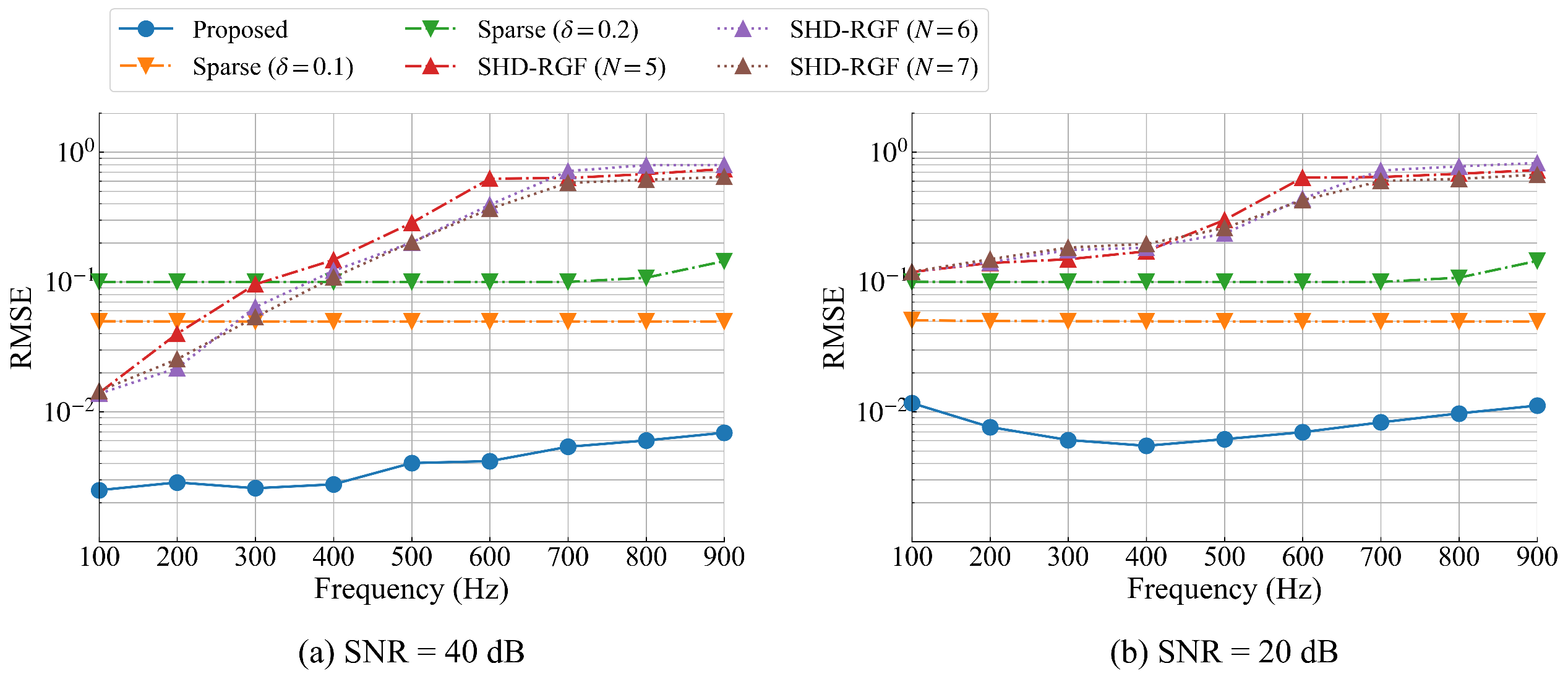}
    \caption{\label{fig:rmse1src}{RMSE as a function of frequency in the case of a single source with (a) an SNR of 40 dB and (b) an SNR of 20 dB. (Color online)}}
\end{figure}
Figure~\ref{fig:rmse1src} shows the RMSE plotted against frequency for the frequency range 100-900 Hz at intervals of 100 Hz. The RMSEs of \textbf{Sparse} were nearly constant at all of the investigated frequencies and SNRs for each $\delta$. Comparing \textbf{Sparse}~($\delta=0.1$) and \textbf{Sparse}~($\delta=0.2$) reveals that a smaller discretization interval resulted in smaller RMSEs, although the RMSEs remained at almost $\delta/2$. Figure~\ref{fig:rmse1src}(a) shows that the RMSEs of \textbf{SHD-RGF} were smaller than those of \textbf{Sparse} at frequencies less than 200 Hz. However, the RMSEs of \textbf{SHD-RGF} increased with increasing frequency because of spatial aliasing. In addition, Fig.~\ref{fig:rmse1src}(b) shows that the RMSEs of \textbf{SHD-RGF} were larger than those of \textbf{Sparse} because noise increased at all of the investigated frequencies. However, \textbf{Proposed} achieved much smaller RMSEs than the other methods at all frequencies under both investigated SNRs. 

\begin{figure}[t]
    \includegraphics[width=\columnwidth]{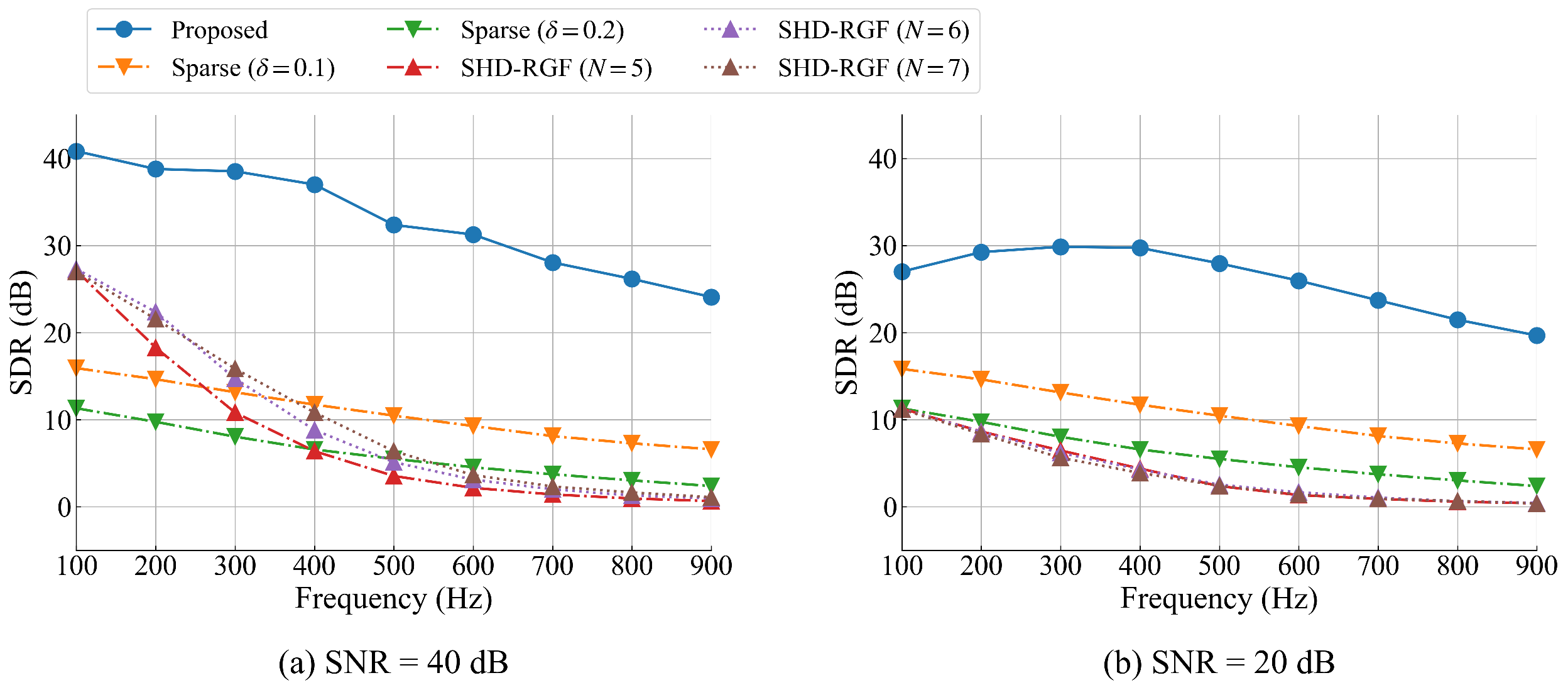}
    \caption{\label{fig:sdr1src}{SDR as a function of frequency in the case of a single source with (a) an SNR of 40 dB and (b) an SNR of 20 dB. (Color online)}}
\end{figure}

Figure~\ref{fig:sdr1src} shows the SDR plotted against frequency for the frequency range 100-900 Hz at intervals of 100 Hz. The SDRs of \textbf{SHD-RGF} were higher than those of \textbf{Sparse} for high SNRs (Fig.~\ref{fig:sdr1src}(a)) and at frequencies less than 200 Hz. The \textbf{Proposed} SDRs were the highest among the SDRs of the investigated methods for all of the considered frequencies and SNRs.

\begin{figure}[t]
    \includegraphics[width=\columnwidth]{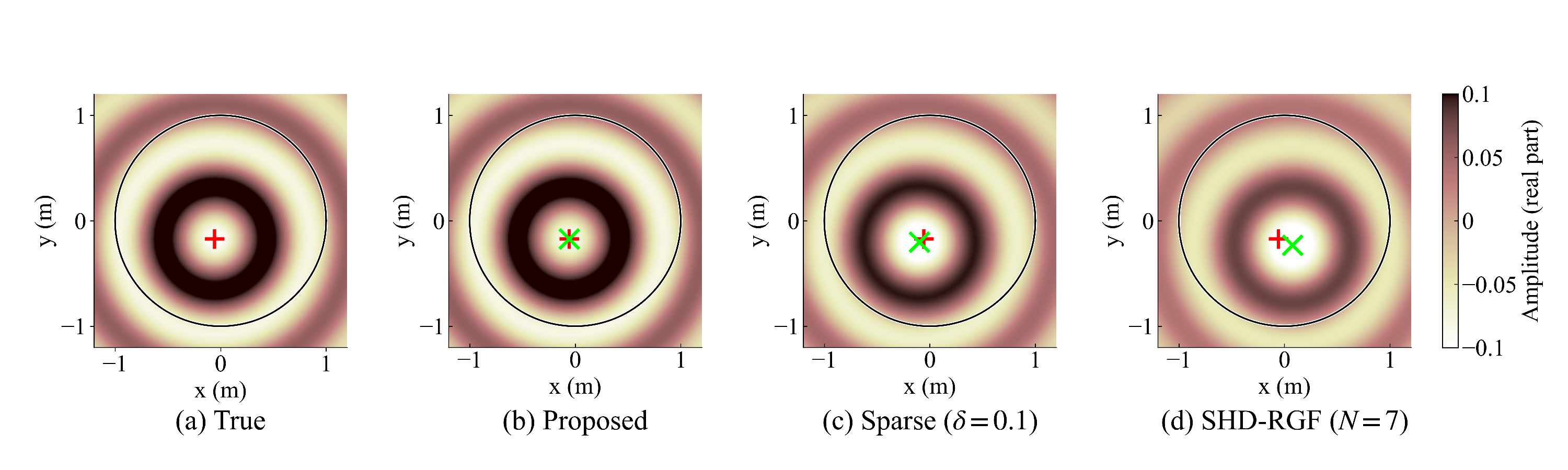}
    \caption{\label{fig:sf1src}{Real part of true and reconstructed sound pressure distribution at 500 Hz on the $x$-$y$ plane in the case of a single source with an SNR of 20 dB. The red and green crosses represent the true and estimated sound source positions, respectively. The black lines represent the sphere where microphones exist. (Color online)}}
\end{figure}

\begin{figure}[t]
    \includegraphics[width=0.8\columnwidth]{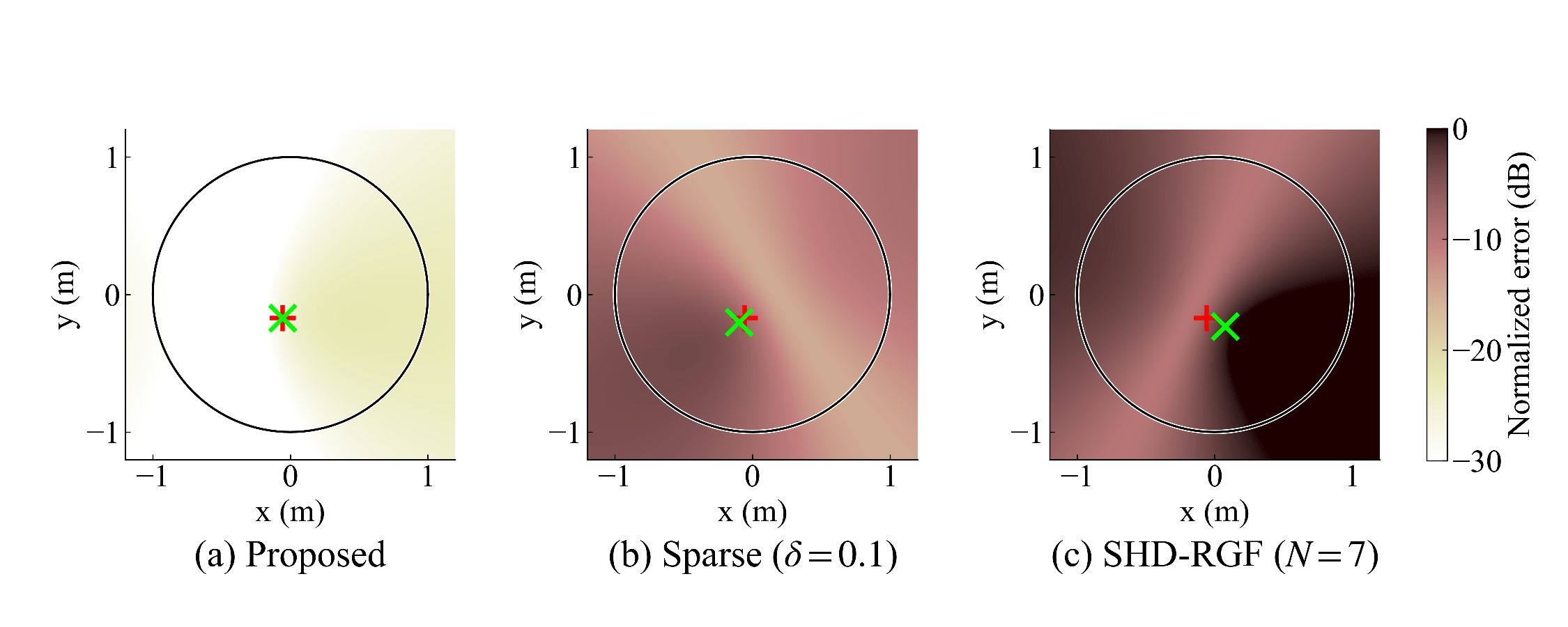}
    \caption{\label{fig:ne1src}{Normalized error distribution at 500 Hz on the $x$-$y$ plane in the case of a single source with an SNR of 20 dB. The SDRs in (a), (b), and (c) were 26.3, 5.8, and 1.8 dB, respectively. (Color online)}}
\end{figure}

Figures~\ref{fig:sf1src} and \ref{fig:ne1src} show the reconstructed sound pressure distribution and the normalized error distribution at 500 Hz on the $x$-$y$ plane for a single source with an SNR of 20 dB. The true position of the source was at $(-0.05,-0.17,-0.45)$, which was chosen randomly from the validation dataset. The amplitude was set to unity. For \textbf{Sparse} and \textbf{SHD-RGF}, only the results of \textbf{Sparse}~($\delta=0.1$) and \textbf{SHD-RGF}~($N=7$) are shown. The red and green crosses correspond to the true and estimated sound source positions, respectively. The black lines represent the sphere where microphones exist. Figure.~\ref{fig:ne1src} shows that \textbf{Proposed} achieved the lowest normalized error distribution among the investigated methods. The SDRs in \textbf{Proposed}, \textbf{Sparse}~($\delta=0.1$), and \textbf{SHD-RGF}~($N=7$) were 26.3, 5.8, and 1.8 dB, respectively.

\subsection{\label{subsec:4:3} Experiments for two sources}
In this experiment, we randomly chose 1,000 validation data from all of the SFS validation datasets which are described in Sec.~\ref{subsec:3:3}. The source signal $\{a_s(k)\}_{s\in\mathcal S}$ was randomized for each condition following $\Re\left[a_s(k)\right]\sim{\mathcal U}(-1,1),\ \Im\left[a_s(k)\right]\sim{\mathcal U}(-1,1)$. The experiments were conducted for SNRs of 40 dB and 20 dB, respectively. The results were averaged for all conditions.

\begin{figure}[t]
    \includegraphics[width=\columnwidth]{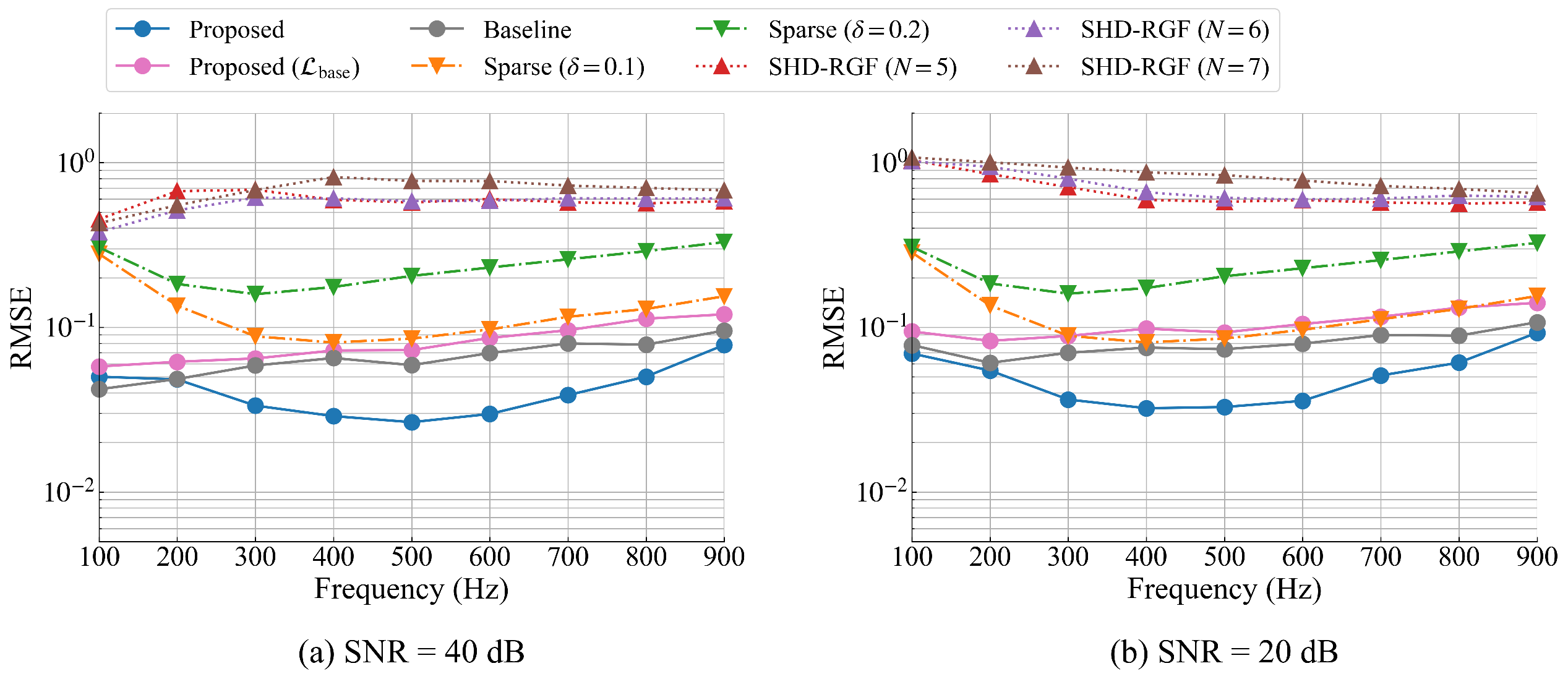}
    \caption{\label{fig:rmse2src}{RMSE as a function of frequency in the case of two sources with (a) an SNR of 40 dB and (b) an SNR of 20 dB. (Color online)}}
\end{figure}

\begin{figure}[t]
    \includegraphics[width=\columnwidth]{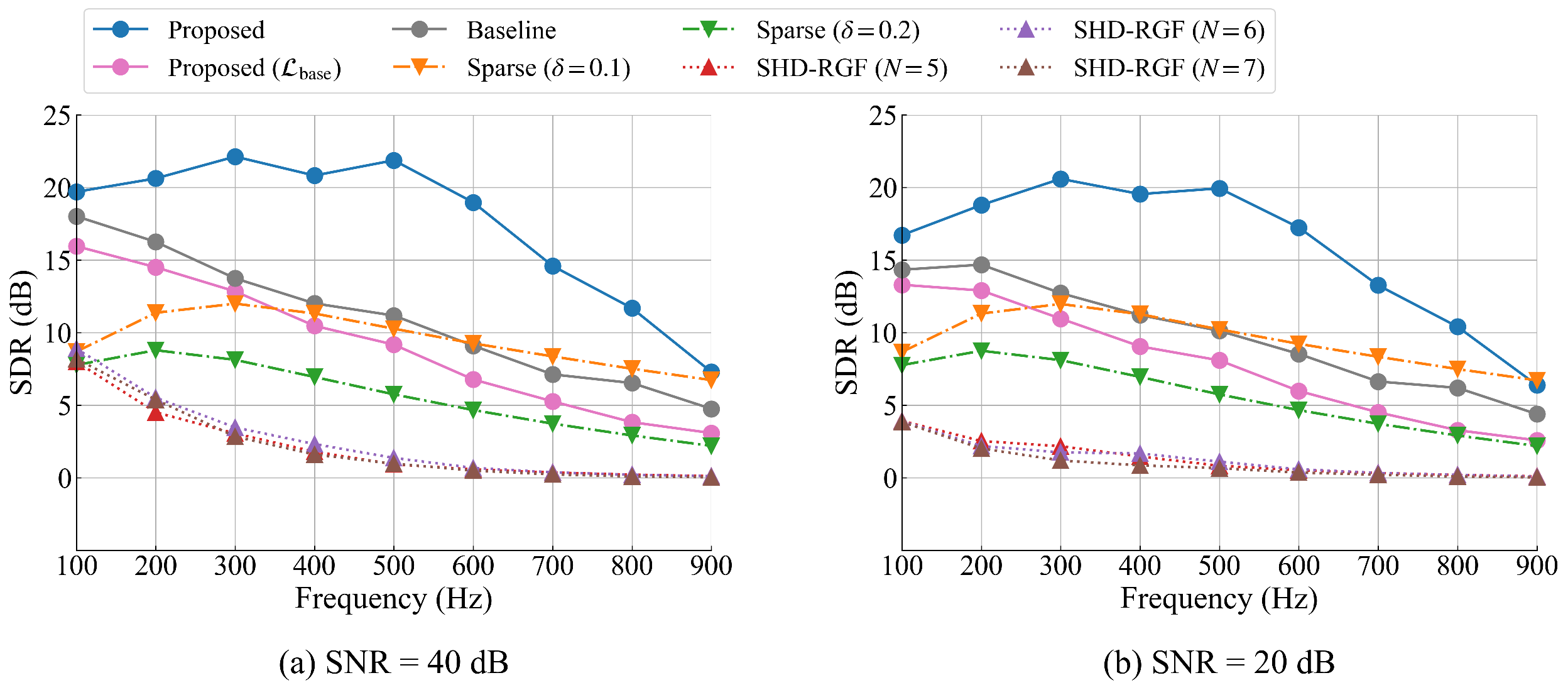}
    \caption{\label{fig:sdr2src}{SDR as a function of frequency in the case of two sources with (a) an SNR of 40 dB and (b) an SNR of 20 dB. (Color online)}}
\end{figure}
Figure~\ref{fig:rmse2src} shows the RMSE plotted against frequency for frequencies ranging from 100 Hz to 900 Hz at intervals of 100 Hz in the case of two sources. In \textbf{Sparse}, the RMSE increased with increasing frequency; this trend differs somewhat from that in the case with a single sound source (Fig.~\ref{fig:rmse1src}). Comparing \textbf{Sparse}~($\delta=0.1$) and \textbf{Sparse}~($\delta=0.2$) reveals that a smaller discretization interval resulted in smaller RMSEs even in the case of two sound sources. Unlike the case of a single sound source shown in Fig.~\ref{fig:rmse1src}, the RMSEs of \textbf{SHD-RGF} were largest at all frequencies under both investigated SNR conditions. Comparing \textbf{Proposed} ($\mathcal{L}_{\rm base}$) and \textbf{Baseline} reveals that the RMSEs of \textbf{Baseline} were smaller than those of \textbf{Proposed} ($\mathcal{L}_{\rm base}$). However, \textbf{Proposed} achieved much smaller RMSEs than the other investigated methods at all frequencies under both SNR conditions, except for 100 Hz with an SNR of 40 dB. These results not only demonstrate the effectiveness of the proposed method but also the effectiveness of the proposed loss function.

Figure~\ref{fig:sdr2src} shows the SDR plotted against frequency for frequencies in the range of 100-900 Hz at intervals of 100 Hz. \textbf{Proposed} achieved the highest SDRs among the investigated methods for all frequencies and SNRs; it also achieved the lowest RMSEs. Remarkably, at 500 Hz, the SDR was more than 10 dB greater than those of the other methods.

\begin{figure}[t]
    \includegraphics[width=\columnwidth]{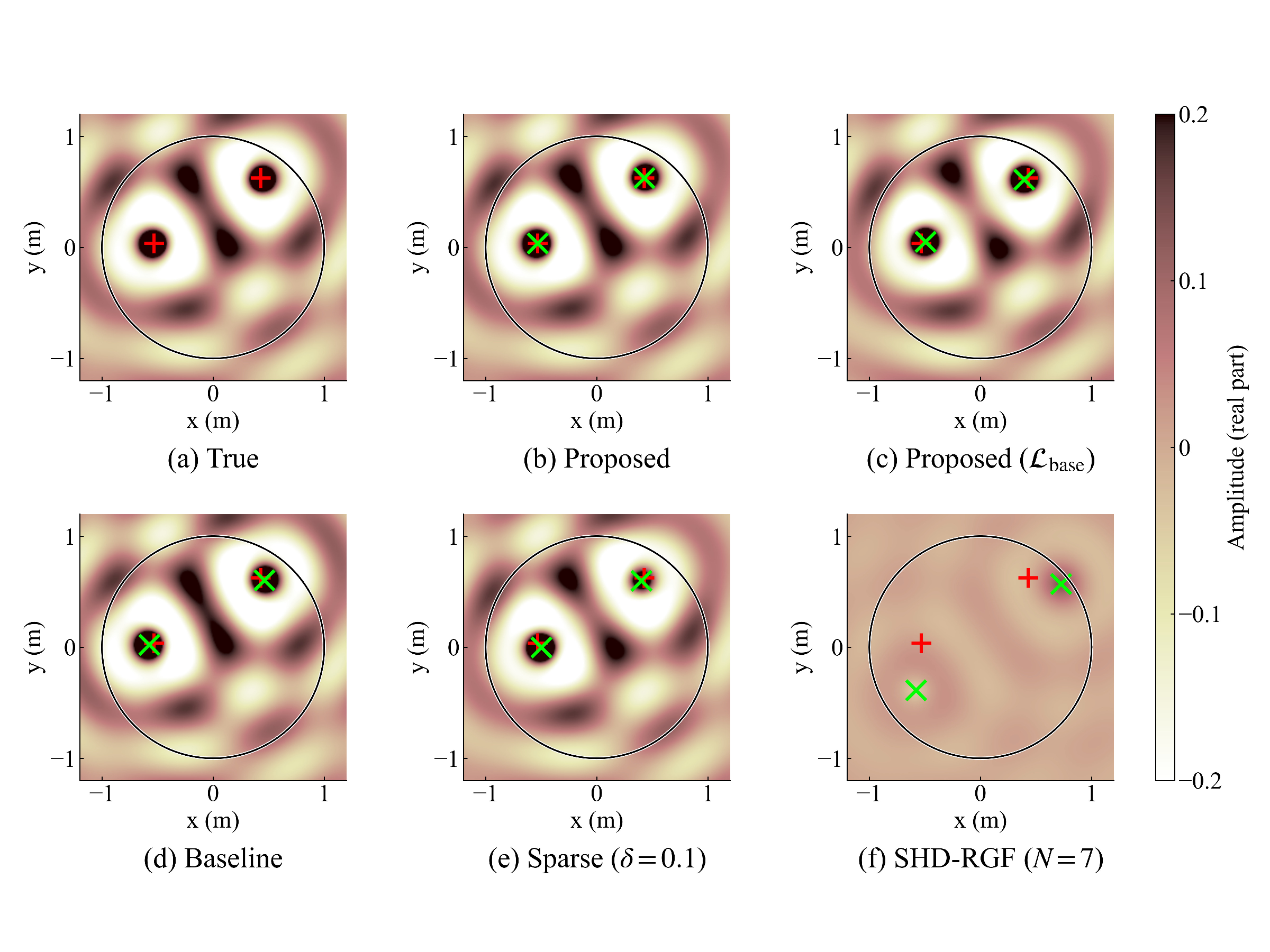}
    \caption{\label{fig:sf2src}{Real part of true and reconstructed sound pressure distribution at 500 Hz on the $x$-$y$ plane in the case of two sources with an SNR of 20 dB. (Color online)}}
\end{figure}

\begin{figure}[t]
    \includegraphics[width=\columnwidth]{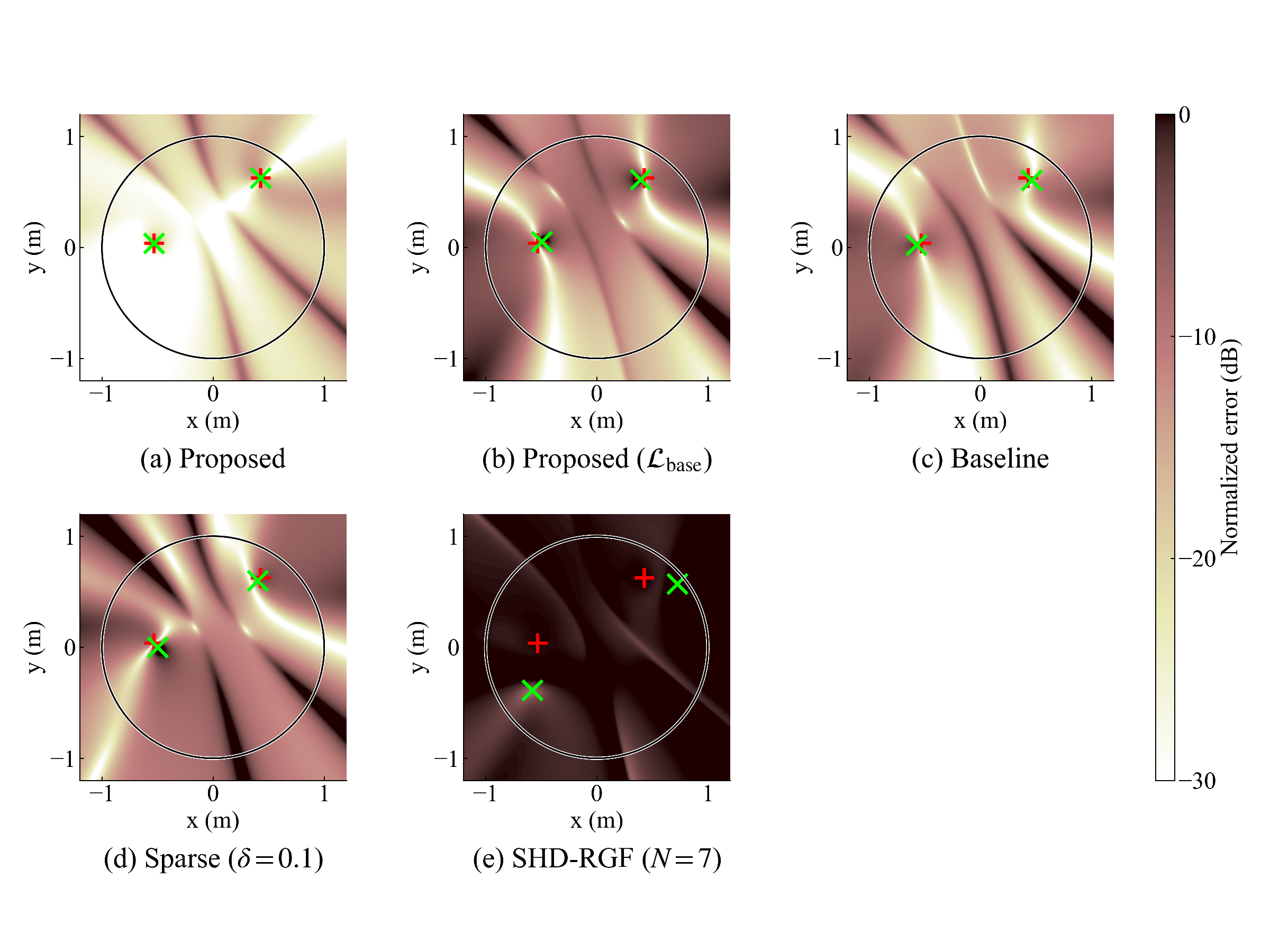}
    \caption{\label{fig:ne2src}{Normalized error distribution at 500 Hz on the $x$-$y$ plane in the case of two sources with an SNR of 20 dB. The SDRs in (a), (b), (c), (d), and (e) were 19.5, 8.0, 11.9, 7.0, and 0.1 dB, respectively. (Color online)}}
\end{figure}

Figures~\ref{fig:sf2src} and \ref{fig:ne2src} show the reconstructed sound pressure distribution and the normalized error distribution at 500 Hz on the $x$-$y$ plane for an SNR of 20 dB. The true positions of the sources were at $(-0.54,\ 0.04,\ -0.04)$ and $(0.44,\ 0.61,\ 0.06)$, which were chosen randomly from the validation dataset. The amplitudes of the sources were set to one. For \textbf{Sparse} and \textbf{SHD-RGF}, only the results of \textbf{Sparse}~($\delta=0.1$) and \textbf{SHD-RGF}~($N=7$) are shown. Figure~\ref{fig:ne2src} shows that \textbf{Proposed} achieved the lowest normalized error distribution among the investigated methods. The SDRs in \textbf{Proposed}, \textbf{Proposed}~($\mathcal{L}_{\rm base}$) , \textbf{Baseline}, \textbf{Sparse}~($\delta=0.1$), and \textbf{SHD-RGF}~($N=7$) were 19.5, 8.0, 11.9, 7.0, and 0.1 dB, respectively.

\clearpage
\section{\label{sec:5} Conclusion}
A neural-network-based method for sound field decomposition was proposed. To reconstruct a sound field in a source-included region, some constraints are necessary to address the ill-posed problem. Conventional methods that use sparsity in the number of sound sources have the disadvantage of determining source-position candidates in advance, which results in a loss of accuracy when sound sources exist at locations other than the candidate locations. In other conventional methods that uses the reciprocity of the transfer function, the accuracy of sound field reconstruction is limited to the spatial Nyquist frequency. To overcome these problems, we proposed two-stage neural networks for sound field decomposition. In the first stage, the sound pressure at microphones is separated into the sound pressure corresponding to each source. In the second stage, the position of each source is localized. For training the first stage, a loss function that explicitly separates measured sound pressure into the sound pressure corresponding to each source was also proposed. Numerical experiments showed that the proposed method that uses the proposed loss function achieved more accurate sound source localization and sound field reconstruction than the investigated conventional methods. Future work will consider the non-anechoic conditions where room reflections exist.

\begin{acknowledgments}
This research was partially supported by JSPS Grants (Nos. JP19H04153 and JP22H00523).
\end{acknowledgments}

\bibliography{sampbib}

\end{document}